# Design of Angular-offset Interstitial-Tube-Assisted Hollow-Core Fibers with Ultrahigh Mode Purity and Ultralow Loss

HAO CHEN[1,2], JIAYI CHEN[1,2], DAWEI GE[3], DONG WANG[3], SHOUFEI GAO[1,2,4], WEI DING[1,2,4], DECHAO ZHANG[3,5], YINGYING WANG[1,2,4,6]

[1] *Guangdong Provincial Key Laboratory of Optical Fibre Sensing and Communication, Institute of Photonics Technology, Jinan University, Guangzhou 510632, China*
[2] *College of Physics & Optoelectronic Engineering, Jinan University, Guangzhou 510632, China*
[3] *Department of Fundamental Network Technology, China Mobile Research Institute, Beijing, China*
[4] *Linfiber Technology (Nantong) Co., Ltd. Jiangsu 226010, China*
[5] *zhangdechao@chinamobile.com*
[6] *wangyy@jnu.edu.cn*

**Abstract:**

Antiresonant hollow-core fibres (AR-HCFs) have recently reached attenuation far below the Rayleigh-scattering limit of silica, but their inherently multimode nature remains a major challenge for practical systems requiring high modal purity. In particular, suppressing higher-order modes (HOMs) at the ~1 dB/m level while maintaining sub-0.1 dB/km fundamental-mode (FM) loss is difficult because conventional filtering strategies rely on tuning nested-tube dimensions, a design freedom that becomes increasingly restricted in the ultralow-loss regime. Here, we propose a new HOM-control mechanism in an interstitial-tube-assisted double nested anti-resonant nodeless fiber (IT-DNANF) by introducing angular offset of the interstitial tubes. Instead of using nested cavities as the primary tuning element, the proposed approach exploits the gap region between adjacent cladding tubes as a leakage-adjacent modal-control interface. Numerical simulations show that the offset increases both FM and HOM losses, but with a substantially stronger sensitivity for HOMs, leading to rapid enhancement of differential modal loss. Furthermore, when the gap-region FM is tuned into phase matching with the core HOM, strong coupling to a high-leakage state is induced, resulting in a pronounced HOM-loss peak. Using the practical criterion of HOM loss > 1 dB/m, we identify optimized IT-DNANF designs that achieve rapid HOM stripping while maintaining FM loss below 0.05 dB/km at 1550 nm. This work establishes angular offset as a physically distinct and manufacturability-friendly degree of freedom for mode purification in ultralow-loss hollow-core fibres.

## 1. Introduction

Antiresonant hollow-core fibres (AR-HCFs) have rapidly evolved into a transformative platform for next-generation optical communications. By guiding light predominantly in air, AR-HCFs fundamentally decouple transmission from the material Rayleigh scattering limit of silica, enabling attenuation below 0.1 dB/km [1] and breaking the long-standing 0.14 dB/km floor of solid-core fibres [2]. Recent progress has demonstrated both record-low attenuation of 0.052 dB/km over 40 km and 0.076 dB/km over 83 km single-draw spans and significant advances in structural scalability through the interstitial-tube-assisted double nested anti-resonant nodeless fiber (IT-DNANF) [3] architecture. In parallel, reduced-diameter triple-nested designs (TNANFs) have pushed confinement loss to 0.25 dB/km level within SMF-compatible packages [4]. These breakthroughs position AR-HCFs as serious contenders for long-haul transmission, data-centre interconnects, and precision sensing systems.

Despite these breakthroughs, a persistent challenge remains: HCFs are inherently multimode structures [5,6]. For applications requiring high-speed coherent transmission [7], high-precision fiber-optic gyroscopes [8,9], or high-beam-quality laser [10], the presence of higher-order modes (HOMs) leads to multi-path interference (MPI) and modal noise, which can severely degrade system performance. Unlike step-index fibres governed by total internal reflection and the V-number criterion, AR-HCF guidance relies on antiresonant reflection and leaky-mode resonance [11,12]. Both the LP01 fundamental mode (FM) and higher-order modes (HOMs) satisfy the antiresonant condition and can propagate in the hollow core. Although HOMs naturally exhibit larger leakage due to their increased grazing angles, this differential loss is typically insufficient to suppress dominant modes such as LP11 over practical fibre lengths.

In the absence of a strict theoretical cutoff, single-mode performance need therefore be defined operationally. According to ITU-T G.650.1 recommendation [13], the cutoff wavelength corresponds to the point where the total power (FM + HOM) exceeds the FM power by less than 0.1 dB. This criterion implies a modal suppression of approximately 19.3 dB. For a standard test length of 22 meters, this level of suppression necessitates an HOM loss level on the order of 1 dB/m. Achieving such rapid HOM stripping while maintaining sub-0.1 dB/km FM loss becomes a stringent and non-trivial design requirement.

Several strategies have been proposed to achieve efficient HOM filtering. One approach introduced geometric asymmetry or enlarged cladding gaps to enhance leakage channels [14], but at the expense of increased FM loss, limiting its applicability for long-haul transmission. A more selective paradigm utilizes phase-matching or resonant coupling [15]. By precisely tuning the dimensions of the cladding air cavities, the cladding air modes can be phase-matched to the core HOMs. This leads to strong coupling at the anti-crossing point, effectively stripping the HOMs from the core into high-leakage states. For instance, the recently reported fourfold truncated double-nested structure (4T-DNANF) achieved a record-high HOM extinction ratio, with an experimental FM loss of 0.13 dB/km and an HOM loss of 6500 dB/km [16]. However, these traditional filtering mechanisms rely heavily on tuning the dimensions of the nested tubes, a design freedom that is increasingly constrained as AR-HCFs approach their confinement-limited regime. In TNANFs, additional tubes are introduced primarily to suppress confinement loss, leaving minimal geometric freedom for HOM-specific tuning. As a consequence, the achievable HOM loss drops to values as low as ~95 dB/km, significantly below the levels required for rapid modal stripping in short fibre lengths [4]. This illustrates a fundamental limitation of relying solely on nested tube dimension engineering for mode purification.

In this work, we focus on the IT-DNANF architecture—a structurally tolerant design that has demonstrated superior draw stability, positioning it as one of the most promising candidates for future large-scale industrial deployment of ultralow-loss HCF. We propose a novel degree of freedom for HOM control: the angular offset of the interstitial tubes. Unlike conventional approaches that rely on adjusting nested tube dimensions, we intentionally exploit the gap region as the primary tuning interface. The gap is spatially closer to the cladding leakage channels, and HOMs, owing to their larger transverse extent and stronger field overlap with the core boundary, exhibit significantly greater interaction with this region. Numerical simulations show that increasing the offset elevates both FM and HOM losses, but with a substantially higher sensitivity for HOMs, leading to a rapid enhancement of the differential loss ratio. More importantly, when the FM in the gap region satisfies the phase-matching condition with the core HOM, a strong coupling leakage channel emerges, causing the HOM loss to peak. This mechanism aligns with the established phase-matching paradigm but shifts the tuning variable from cavity dimension to a spatial positioning dimension, providing a new structural pathway for enhancing mode purity. Based on the criterion of HOM loss > 1 dB/m, we present optimized IT-DNANF designs that achieve rapid HOM stripping while maintaining an ultra-low FM loss (< 0.05 dB/km). This approach offers a manufacturability-friendly route for the design of high-mode-purity, ultra-low-loss hollow-core fibers for future optical communication networks.

**2. Fiber design**

The fiber structure under consideration is IT-DNANF in which interstitial tubes (IT) are placed in the gaps between adjacent cladding tubes. The initial structural parameters were chosen by balancing low FM loss, geometric manufacturability, and the possibility of activating the gap region as a HOM-filtering interface. The core radius was fixed at 14 μm and the wall thickness at 1.1 μm. Under these conditions, an inter-tube gap of 8–10 μm was selected as the main design range, since this interval reduces the likelihood of mid-draw contact (MDC) [17] while still allowing the gap region to remain large enough for leakage control. Because the IT is used to regulate the effective gap area, its radius must be carefully chosen. If the IT is too small, it cannot sufficiently shape the gap region and cannot effectively prevent unwanted energy leakage, which leads to increased FM loss. If it is too large, it can over-block the HOM leakage pathway and suppress the very extraction mechanism that is being targeted. After initial screening, IT radii between 4 μm and 6 μm were identified as the most suitable range for maintaining low FM loss without completely blocking HOM leakage. Therefore, three representative cases with IT radii of 4 μm, 5 μm, and 6 μm were studied in detail. All simulations in this study were performed using the finite element method (FEM) with an optimized perfectly matched layer (PML).

We first examine whether the conventional HOM-filtering route based on the nested-tube air regions can provide sufficiently strong HOM suppression in the present IT-DNANF platform. In this design, each IT is aligned with the center of the adjacent cladding gap. Figure 1a illustrates the FM and HOM loss at a 1550 nm wavelength for the IT-centered structure (Fig.1d) with varying IT radii. Structures utilizing different air cavities within the nested tubes for HOM filtering were simulated. Fig.1b shows results for HOM filtering via the region between the medium and small tubes (Region 1), while Fig.1c corresponds to the region between the large and medium tubes (Region 2). Fig.1e & f illustrates the phase-matching conditions between the core HOM and the specific tuning areas' (Region 1 / Region 2) FMs. As shown in Fig. 1a, when optimization is focused solely on FM loss, then HOM loss usually remains between ~1 and 10 dB/km (e.g., points ①, ②). As the gap widens and IT's size decreases, both FM and HOM loss gradually increases, with HOM loss reaching ~70 dB/km at the expense of the increased FM loss of 0.43 dB/km (points ③ and ④). Under phase-matching conditions between FM air mode in Region 1 and HOM in the air core (Fig.1e), HOM loss reaches the order of ~$10^2$ dB/km, while FM loss also increases to 0.43 dB/km (e.g., points ⑤ and ⑥ in Fig. b). Similarly, utilizing phase matching in Region 2 yields FM and HOM losses on the order of ~$10^{-2}$ dB/km and $10^2$ dB/km (points ⑦ and ⑧ in Fig. 1c). In all cases, HOM loss fails to meet the 1 dB/m threshold for effective stripping. This is because the nested-region route is not structurally adjacent enough to the strongest cladding leakage boundary to generate fast HOM removal. Furthermore, as structural complexity increases and more nested layers are introduced to reduce FM loss, the number of geometrically useful and independently tunable regions for selective HOM extraction can actually decrease. In this sense, conventional nested-region tuning becomes increasingly constrained precisely in the regime where ultra-low FM loss is most desired. It should be noted that these results are numerical estimates obtained within the present simulation framework. In practice, the actual HOM loss in fabricated fibers may become higher than simulated values, and filtering through the conventional nested regions may still be acceptable in applications that do not require extremely strong modal purity. Only under the stricter criterion adopted here, the nested-region route is not sufficient. This inadequacy directly motivates the search for a stronger tuning interface—one that is geometrically closer to the leakage boundary and can provide a more efficient coupling-mediated escape path for the HOM.

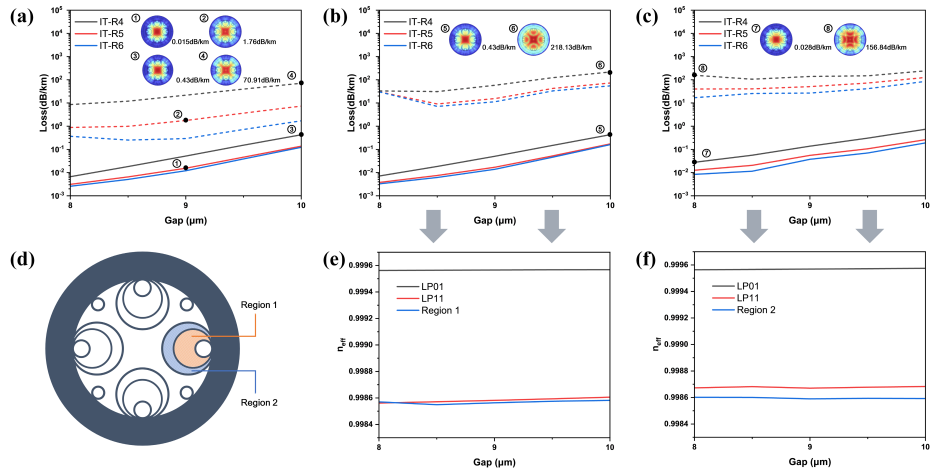

Figure 1. Conventional nested-region HOM filtering in the IT-centered DNANF structure. (a–c) FEM-simulated confinement loss of LP01 (solid lines) and LP11 (dashed lines) as a function of inter-tube gap for three interstitial-tube radii (IT-R): 4 µm (black), 5 µm (red), and 6 µm (blue). Specifically: (a) non-phase matching structure; (b) HOM filtering using the region between the medium and small nested tubes (Region 2). (c) HOM filtering using the region between the large and medium nested tubes (Region 1). (d) Schematic drawing of the simulated structure indicating the two tuning regions. (e–f) Effective refractive indices of $LP_{01}$ (black), $LP_{11}$ (red), and the corresponding region modes (blue) illustrating the phase-matching conditions for Region 2 (e) and Region 1.

The gap region between adjacent cladding tubes is normally not treated as a useful HOM-filtering region in conventional DNANF structures because it is too large and does not naturally satisfy the size condition for phase matching with the core HOM. In IT-DNANF, however, the introduction of an IT changes this situation. By partially partitioning the gap region, the IT reshapes the effective air area and makes the gap region structurally accessible as a modal-control interface. Yet in the symmetric centered configuration, the IT divides the original gap region into two smaller sub-regions. Because these two halves do not satisfy the required size condition for strong phase-matching-based coupling to the core HOM, the gap region in the baseline state is present but not yet an optimized extraction interface. The next step is therefore to ask whether the gap region can be activated more deliberately.

To do this, we introduce angular offset of the IT. Starting from the centered 0° state, the IT is rotated counter-clockwise relative to the fiber center. This breaks the symmetry of the local gap geometry and enlarges the effective air area on one side of the gap while compressing it on the other. The angular-offset simulations were carried out for structures with inter-tube gaps of 8–9 µm and interstitial-tube radii of 4–6 µm. Larger gap values were avoided because they led to unacceptably high FM loss. Figures 2(a)–2(c) illustrate the cross-sectional evolution of the offset configurations for the three IT radii. Figures 2(d)–2(f) show the corresponding FM and HOM losses at 1550 nm as functions of inter-tube gap distance and offset angle. A clear pattern emerges. As the offset angle increases, the FM loss changes steadily and shows an overall increasing trend. The HOM loss also generally increases, but its evolution is not simply monotonic. In some cases, the HOM loss initially rises and then falls. In others, it exhibits a pronounced peak over a limited angular interval. These distinct features indicate that more than one mechanism is at work. For small offsets, the dominant contribution is the progressive alteration of the leakage environment. However, as the offset continues to increase, a more selective interaction becomes possible. The effective index of the gap-region FM moves closer to that of the core HOM, and the system approaches a phase-matching condition. When this happens, the HOM loss increases dramatically over a narrow angular interval, producing the observed peaks.

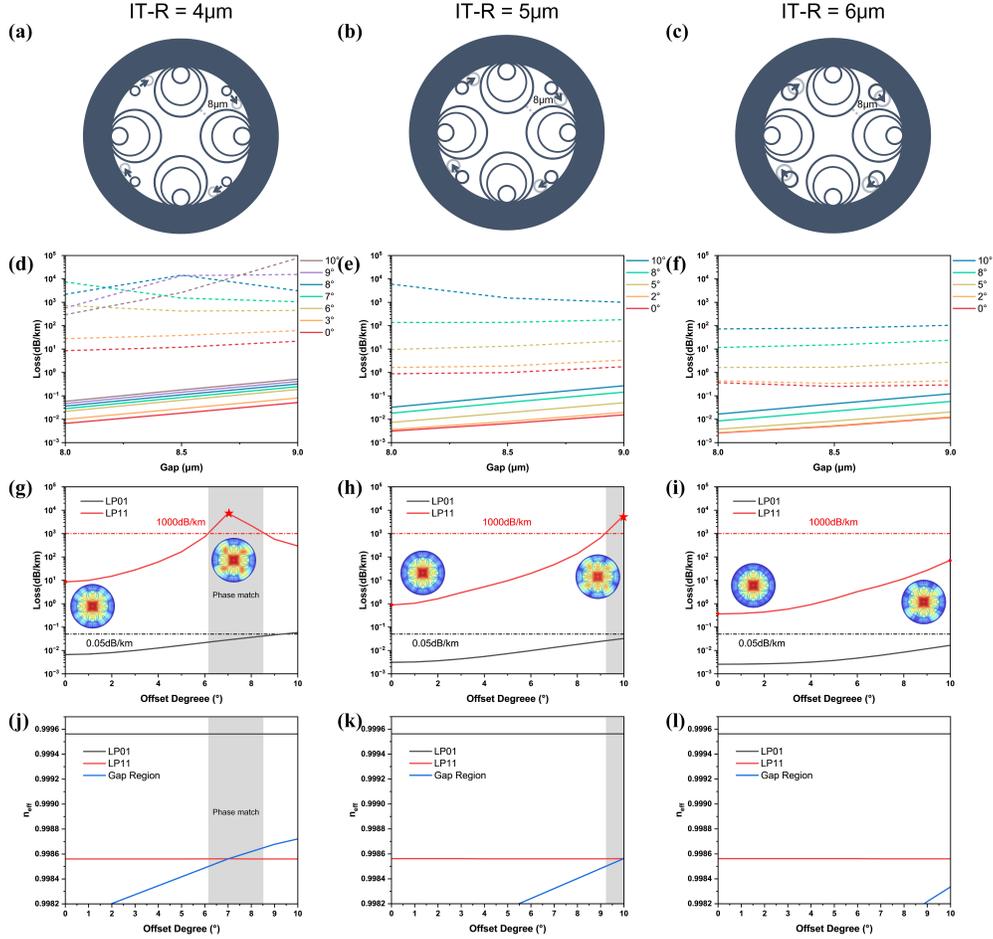

Figure 2. Effect of angular offset of the interstitial tube on modal losses. (a–c) Cross-sectional geometries of IT-DNANF with angular offset for IT radii of 4 μm, 5 μm, and 6 μm, respectively. (d–f) FEM-calculated confinement loss of $LP_{01}$ (solid lines) and $LP_{11}$ (dashed lines) at 1550 nm as a function of offset angle for the corresponding structures. (g–i) Modal losses versus offset angle for an inter-tube gap of 8 μm; grey shaded regions indicate the phase-matching intervals where HOM loss is enhanced. Insets show the modal field distributions for representative configurations. (j–l) Effective refractive indices of $LP_{01}$ (black), $LP_{11}$ (red), and the gap-region modes (blue) as a function of offset angle, illustrating the phase-matching mechanism.

Figures 2(g)–2(i) make this point more directly by showing the modal losses at an inter-tube gap distance of 8 μm as a function of offset angle. Grey-shaded regions identify the angular ranges where phase matching is approached. In the 4 μm interstitial-tube case, a strong HOM-loss peak appears near a 7° offset angle. In the 5 μm case, a similar peak occurs near 10°. For the 6 μm case, the enhancement is weaker, indicating that no phase matching condition has been reached. The corresponding effective-index evolution, shown in Figures 2(j)–2(l), confirms this interpretation. As the offset angle changes, the effective index of the gap-region FM shifts significantly, while the core FM and HOM index remains relatively stable. The sharp HOM-loss enhancement occurs precisely where the core HOM and the gap-region FM

approach one another in effective index. This is strong evidence that the dominant mechanism is not generic asymmetry-induced leakage but coupling mediated by phase matching.

For the optimized points marked in Figures 2(g) and 2(h), the HOM losses reach approximately 7.4 dB/m for the 4 μm IT radius at a 7° offset angle and approximately 5.9 dB/m for the 5 μm radius at a 10° offset angle. In both cases, the FM loss remains below 0.05 dB/km at 1550 nm. These values represent a striking improvement over the conventional nested-region route. The physical reason is straightforward but important. Once the gap-region FM becomes phase matched to the core HOM, the gap region acts as a leakage-efficient extraction interface because it is structurally close to the cladding boundary. Compared with the nested cavities, it provides a shorter and more direct route for power transfer from the HOM to the external leakage channel. The offset angle serves as the control parameter that enables this extraction channel to be activated selectively. In this sense, angular offset is not merely another geometric tuning variable. It is a means of turning the gap region into a resonantly accessible HOM sink. At the same time, the FM remains protected by its weaker overlap with the gap region and by the fact that no comparable resonant interaction is established for the FM in the optimized designs. This explains why the differential loss between FM and HOM can be dramatically increased without destroying the ultra-low-loss guiding behavior of the FM.

To assess the practicality of the optimized designs, we next examined the broadband loss behavior of the two best-performing structures under both straight and bent conditions. In Fig. 3, Structure A corresponds to the 4 μm IT- radius with a 7° offset angle, while Structure B corresponds to the 5 μm radius with a 10° offset angle. Under straight conditions, both structures maintain strong HOM suppression across a broad spectral range in the O to C band while preserving low FM confinement loss. The peak HOM losses are approximately 8.3 dB/m for Structure A and 6.5 dB/m for Structure B, whereas the FM losses remain around 0.041 dB/km and 0.046 dB/km, respectively. These results show that the optimized gap-region route remains effective over a practically useful spectral window.

The effect of bending introduces a more nuanced picture. When the optimized structures are bent, the FM loss remains comparatively stable, indicating that the basic antiresonant guiding mechanism of the FM is not strongly disrupted. In contrast, the HOM suppression becomes more sensitive to bending. This behavior can be understood in terms of the same physical mechanism that produced the strong HOM filtering under straight conditions. Because the HOM attenuation relies on phase matching between the core HOM and the gap-region FM, any perturbation that shifts the effective indices of these modes can weaken the coupling. Bending changes the local modal environment and therefore perturbs the phase mismatch condition. As a result, the HOM-loss peak can be reduced or shifted. This does not undermine the validity of the gap-region filtering concept; rather, it reveals its operating boundary. The same resonant sensitivity that produces strong HOM stripping under optimized straight conditions also makes the HOM attenuation more vulnerable to external perturbations. In other words, the price of using a stronger and more selective extraction channel is an increased sensitivity of the HOM-filtering state to bending-induced modal detuning. This trade-off has practical implications. For applications that require extremely high modal purity and operate under relatively straight or gently perturbed fiber conditions, the proposed gap-region route is highly attractive. It provides much stronger HOM suppression than conventional nested-region tuning while preserving ultra-low FM loss. For applications where bending robustness is the dominant concern, more moderate filtering strategies may still be preferable, including routes that sacrifice some stripping strength in exchange for a less phase-sensitive modal response. The important point is that the present work does not claim a universally superior filtering solution for all environments. Instead, it establishes a new and physically distinct design route whose advantages are especially pronounced in the straight-fiber or weak-bending regime.

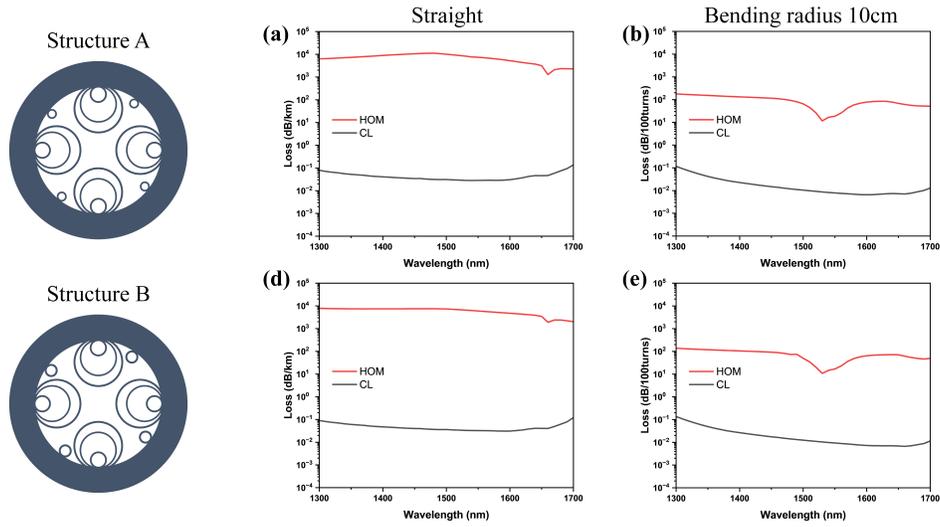

Figure 3. Broadband loss characteristics of the optimized IT-DNANF structures. (a,c) Confinement loss spectra of LP$_{01}$ and LP$_{11}$ for Structure A (IT-R = 4 µm, offset = 7°) and Structure B (IT-R = 5 µm, offset = 10°) under straight fiber conditions. (b,d) Corresponding loss spectra under a bending radius of 10 cm.

## 3. Conclusion

In summary, this work proposes and numerically demonstrates a new HOM-filtering strategy in IT-DNANF. The key idea is to exploit the gap region between adjacent cladding tubes as a primary tuning interface for HOM extraction. In conventional nested antiresonant fibers, this region is generally too large to satisfy the phase-matching condition required for selective HOM filtering and is therefore not normally used for modal control. By introducing ITs, the effective gap area can be reshaped and partially partitioned, making the gap region structurally accessible as a modal-engineering interface. The optimized structures identified in this work achieve HOM losses of approximately 7.4 dB/m and 5.9 dB/m at 1550 nm for 4 µm and 5 µm interstitial-tube radii, respectively, while maintaining FM loss below 0.05 dB/km. Broadband analysis further shows that strong HOM suppression persists across the O to C band under straight conditions. Under bending, the FM remains relatively stable, whereas HOM filtering becomes more sensitive because bending perturbs the phase-matching condition. This reveals a practical trade-off between maximum modal purity and bend robustness.

Overall, the results demonstrate that the gap region in IT-DNANF can function as an efficient and physically distinct HOM-filtering interface. Angular offset emerges as a powerful control parameter that enables phase-matching-assisted extraction of HOMs without sacrificing the ultra-low-loss character of the FM. This opens a new pathway for mode engineering in AR-HCF and provides a promising strategy for achieving high modal purity in future ultra-low-loss hollow-core fiber systems.

## 4. Data availability statements

Data underlying the results presented in this paper are not publicly available at this time but may be obtained from the authors upon reasonable request.

## 5. Fundings

This work was supported by the Scientific Research Innovation Capability Support Project for Young Faculty (ZYGXQNJSKYCXNLZCXM-I18).